# Modification of the hybridization gap by twisted stacking of quintuple layers in a three dimensional topological insulator thin film

Changyuan Zhou[1], Dezhi Song [1], Yeping Jiang[1]* , Jun Zhang [1]

[1] Key Laboratory of Polar Materials and Devices, Department of Electronic, School of Physics and Electronic Science, East China Normal University, Shanghai 200241, China

Twisting the stacking of layered materials leads to rich new physics. A three dimensional (3D) topological insulator film host two dimensional gapless Dirac electrons on top and bottom surfaces, which, when the film is below some critical thickness, will hybridize and open a gap in the surface state structure. The hybridization gap can be tuned by various parameters such as film thickness, inversion symmetry, etc. according to the literature. The 3D strong topological insulator Bi(Sb)Se(Te) family have layered structures composed of quintuple layers (QL) stacked together by van der Waals interaction. Here we successfully grow twistedly-stacked $Sb_2Te_3$ QLs and investigate the effect of twist angels on the hybridization gaps below the thickness limit. We find that the hybridization gap can be tuned for films of three QLs, which might lead to quantum spin Hall states. Signatures of gap-closing are found in 3-QL films. The successful in-situ application of this approach opening a new route to search for exotic physics in topological insulators.

* Corresponding authors.
Email: ypjiang@clpm.ecnu.edu.cn

The class of three dimensional (3D) strong topological insulators (TIs) such as $Bi_2Se_3$, $Bi_2Te_3$, and $Sb_2Te_3$ host single Dirac cone in the surface state structure on the surface due to the non-trivial bulk band topology[1-7]. In the thin film form, the two dimensional (2D) massless Dirac electrons on the top and bottom surfaces will hybridize and open a gap below the critical thickness[3,8-12]. These materials have layered structure composed of quintuple layers (QL) of atoms. There is only van der Waals interaction between the QLs so that the film thickness is highly controllable. Theoretically, the gap shows oscillatory behavior between an inverted and a normal gap parity with film thickness near the thickness limit[10], which provides a possibility of realizing the quantum spin Hall effect (QSH) in the ultra-thin 3D TIs. In addition, the gap value or parity can be tuned by inversion-symmetry-broken parameters such as electric fields normal to the surface[13-15]. The gap-opening behavior was investigated theoretically and experimentally in ultra-thin $Bi_2Se_3$ and $Sb_2Te_3$ films[3,8]. Nonetheless, up to now, QSH has not been realized in 3D TI systems.

Experimentally, materials with layered structures can be cleaved into a-few-layer or even into monolayer films, which can then be stacked together in a controllable manner to get twisted structures with different angles. In graphene the twisted structures at some magic angles drive the electronics into the superconducting or even the strongly-correlated regime[16]. Here we present an in-situ approach to obtain twisted structures of TI QLs. By molecular beam epitaxial and fine-tuning the growth parameters, we obtain structures with different twist angles between QLs in $Sb_2Te_3$ films below the thickness limit. By scanning tunneling microscope (STM) and spectroscope (STS), we find that the hybridization gap is highly tunable by constructing QLs with different twist angles. We obtain a much wider range of gap-tuning compared to the electric field approach, thus providing a new approach for the search of new QSH systems or even new exotic physics beyond the TI regime in this material.

Molecular beam epitaxy (MBE), mostly known as a tool for the growth of ultra-thin film to achieve higher quality, is seldom conducted in a way by using its another inherent feature. That is, the growth dynamics of films by MBE is highly tunable to achieve non-equilibrium growth by changing the crucial parameters such as substrate roughness, lattice mismatch, substrate temperature, flux and flux ratio, etc. While the MBE growth of high quality films of Bi(Sb)Se(Te) family is well established and now routine[17-21], the attempt to extremize the non-equilibrium growth conditions to get films with structures much different from the equilibrium and normal states is rare.

The thickness limit for $Sb_2Te_3$ being a 3D TI is around 4 QLs, below which a hybridization gap opens in the surface states[3]. In this study, we mainly focus on films below 4 QLs. The 3-QL film is most important because of its small intrinsic hybridization gap (~ 60 meV on graphene)[3]. We grow ultra-thin $Sb_2Te_3$ films with a total coverage of 1-3 QLs on $SrTiO_3$(111) substrate in a two-step manner. The substrate surface has a roughness of ~ 200 pÅ (peak to peak) and terrace widths of ~ 80 nm after annealed in vacuum prior to the growth. In normal conditions, such surface is relatively rough for the epitaxial growth. In the case of Bi(Sb)Se(Te) family, because of the QL structures[6], the interaction between films and under-lying substrate is van der Waals in nature if the substrate is inert. We find that the 1-QL film on the STO(111) substrate is freestanding and there is no interfacial buffer layer. In the two-step procedure, the substrate temperature is kept between 200 °C and 220 °C. We first grow a 1-QL film at a relatively low temperature and do the second-step deposition at a higher temperature without post annealing with a total coverage of about 2-3 QLs. Figure 1a shows the STM topographic image of a 3-QL $Sb_2Te_3$ film. The terraces in the figure are from the substrate (step height ~ 2.3 Å), and the film is continuous across the steps of substrate as observed from high resolution images (not shown). From the topographic image we can see some periodic patterns with different periodicities, which are more clearly shown in the differential image (right image in figure 1a).

These periodic patterns are moire patterns formed in the presence of twist angles between adjacent QLs. To clarify their origin in the 3-QL film, we also grow a film with a coverage of ~2.6 QLs. As shown in figure 1c, the 2.6-QL film composes of a complete 2-QL region in addition to some 3-QL patches. By inspecting the moire patterns across the steps between $2^{nd}$ and $3^{rd}$ QLs (not shown), we find that all the moire patterns are from the first two QLs. That is, the patterns on the $3^{rd}$ QL are from the bottom two QLs.

In $Sb_2Te_3$ films, the clover-shaped defect pattern can be used to identify the crystal direction of the material, with the tips of the clover pointing to the $\overline{\Gamma}-\overline{M}$ directions[17]. As shown in figure 1b, there are two kinds of domain boundaries in $Sb_2Te_3$ films, DW1 and DW2. DW1 (dashed-line-like ones) have small mismatch angles between domains across the boundaries and the underlying crystal structure is continuous. On the contrary, the appearance of DW2 indicates a very large mismatch in crystal orientations between domains. In this case, the lattices across such boundaries can no longer match together and true domain boundaries form.

By inspecting the defects, most domains in the 2$^{nd}$ QL are found to be aligned approximately in the same direction. For the 3$^{rd}$ QL, there are 4 dominant directions with respect to the 2$^{nd}$ QL, perpendicular or parallel to the 2$^{nd}$ QL as indicated in figure 1c. In this four configurations between 2$^{nd}$ and 3$^{rd}$ QLs, there is no clear moire pattern, probably because the period is small and the spatial modulation in the electronic states is weak when the twist angle is large.

So the periodic patterns are moire patterns formed in the presence of twist angles between the first two QLs. The presence of twist-angle is caused by different growth dynamics between 1$^{st}$ and 2$^{nd}$ QLs (or 3$^{rd}$ QL) in the presence of different underlying platforms for the growth. First, a rough substrate leads to reduced mobility of Sb and Te atoms or molecules bombarded onto the surface. The reduced substrate temperature during growth further magnifies this effect, leading to an increased nucleation density. Second, because of the large lattice mismatch between $Sb_2Te_3$ and STO(111) surface, domains originated from different nucleation centers normally have different crystal orientations. The growth of 2$^{nd}$ and 3$^{rd}$ QL during the second step basically flows a layer-by-layer mode. In this mode, nucleation for the next QL will not happen until the present QL is nearly complete. This is clearly seen in figure 1a, where the nucleation of the 4$^{th}$ QL appears on a domain boundary of 3$^{rd}$ QL which is nearly complete. Furthermore, the step flow mode normally leads to a QL without domains, contrary to our observation. In the differential image of figure 1c, we clearly see domains in the 2$^{nd}$ QL. The domains with DW1 are probably caused by the local twist of the 2$^{nd}$ QL in the presence of randomly-orientated domains in the 1$^{st}$ QL. In figure 1c, we can see in the 2$^{nd}$ QL domains with different moire patterns and even different moire patterns in a single domain.

Thus, the structures of twisted QLs in our ultra-thin $Sb_2Te_3$ film is concluded as follows. The domains in the 1st QL are small and along different directions. The domains in the 2nd QL are larger and along approximately the same direction. In the 3rd QL, the domains are aligned perpendicularly to or parallel to the 2nd QL. All these inevitably lead to the situation sketched in figure 1d, indicating the formation of regions with random twist-angles between the first two $Sb_2Te_3$ QLs and four configurations between the 2nd and 3rd QLs.

A simple simulation of moire patterns is shown in figure 2. There are three different positions in the unit-cell of $Sb_2Te_3$ QL. Figure 2a shows the moire pattern corresponding to an angle of ~5°. The periods coming from different angles can be fitted by a simple relation $A = a/\sin(\theta/2)$, where $\theta$ is the twist-angle. By such relation, we then can assign twist-angles to the first two QLs. In addition, the moire patterns on the 3rd QL in figure 1a are all coming from those between 1st and 2nd QLs. Thus we can do statistics on the moire patterns between the underlying 1st and 2nd QLs even in the presence of the 3rd QL.

When a 3D TI is reduced to some critical thickness, a hybridization gap forms in the surface state structure (figure 3a)[9,10]. For the 1-QL film, we can see a clear gap-like structure in the STS. The gap is measured to be about 1000 meV (figure 3b), much larger than that grown on the graphene substrate (~ 700 meV) [3] probably because of the strain in the film grown on the STO substrate. Intuitively, when there is a twist between QLs, the overlap of electron wavefunctions in the van-der-Waals gap between QLs changes. The hybridization between top and bottom surfaces might change accordingly. Thus, by measuring the STS on twisted 2-QL and 3-QL films, we can investigate the effect of twist-angle on the hybridization gap.

For the 2-QL film, the gap indeed varies from domain to domain. The gap is solid with sharp edges (figure 3c) and are uniform inside the domains, ruling out other extrinsic effects. The gap values range from ~180 meV to ~230 meV. For the 2-QL film on graphene, the gap is ~230 meV[3], which can be taken as the intrinsic value because the weak interaction between the film and the graphene substrate. In figure 4a, we summarize the values of measured hybridization gaps of differently-twisted 2-QL and 3-QL films. Nonetheless, for 2-QL film, there is no clear angle-dependent pattern. There is probably another factor, such as the strain effect, coming into play dominantly in this situation.

The 3-QL film is our focus here. 60 meV is the gap value of the 3-QL film grown on the graphene substrate, which is taken as the intrinsic value for the 3-QL film of this material[3]. It is more easy to modulate a small gap to achieve a parity change indicated by the gap-closing phenomenon. So we try to investigate the possible topological phase transition in the 3-QL film. We describe the twist configuration by two angles in the form α+β, where α is the twist angle between the 1$^{st}$ and 2$^{nd}$ QLs and β is that between the upper two QLs. There is also no clear dependence of gap values on the twist-angle α between the 1$^{st}$ and 2$^{nd}$ QLs. But the gap values can be divided into two categories by the angle β (the lower two shaded regions in figure 4a). Those gaps around 100 meV are all measured on domains where the 3$^{rd}$ QL is aligned anti-parallel to the 2$^{nd}$ QL (β=180°). Those around 60 meV are measured on domains where the upper 2 QLs are in other three configurations, that is, the two perpendicular directions and the parallel directions (β=0°, ±90°). So the anti-parallel stacking between the 2$^{nd}$ and 3$^{rd}$ QLs plays a dominant role here. This kind of stacking is shown in figure 4b. For the α+β (0°, ±90°) configurations, the gaps range from 35 meV to 70 meV, fluctuating around the intrinsic value of 60 meV. For the α+β (180°) configurations, the gaps vary from ~80 meV to ~110 meV, deviating much from the intrinsic one.

So, in the α+β (180°) cases, we always see large gaps around 100 meV. This means that the 3-QL film of Sb$_2$Te$_3$ is driven into a large-gap system compared to the intrinsic situation in this case. On the contrary, when β = 0°or ±90°, the gaps keep small and are highly tunable and may reach zero in the presence of other parameters such as the angle α and possible strains in the film. Figures 4c and 4d show some typical spectra in two different energy ranges with different gap values. Besides the regions with solid gaps in STS, we do see gapless features (light-blue curves in figure 4d) in some domains belonging to the α+β (0°, ±90°) cases. These possible gapless states are not plotted in the statistics (figure 4a) because the bare STS is not the solid evidence.

Indeed, in the 3-quintuple-layer (QL) structures of $Sb_2Te_3$ films, the only clear conclusion we can draw from the data is that the antiparallel stacking between the QLs will drive the system into the large-gap region. We don't see any regular pattern on how the small twist angle between the first-two QLs changes the hybridization gap. Although the gap does change from domain to domain with different twist angles, there seems no correspondence between them. One possibility is that there may be strain in the film, because the gap (~1000 meV) of the 1-QL film is much larger than that (~700 meV) grown on the graphene substrate. The strain, if there exists, is uniform inside one particular domain because the gap shows no variation inside the domain. Another possibility is that except for the twist there might exist another mismatch between the lattices of the first two QLs, for example, the 2nd QL might slide a little bit away from the thermodynamic equilibrium position. That is, there might be translational mismatch between the two QLs besides the rotational one. The 1$^{st}$ QL is probably considerably strained on the substrate, while the interaction between the QLs is much weaker. This implies a relatively free registration of the 2$^{nd}$-QL lattice with respect to that of the 1$^{st}$ one. Currently we don't have enough evidence to support this scenario.

According to the literature, the hybridization gap can be modulated by a perpendicular electric field[14,22]. But up to now no clear evidence of gap-closing signature has been discovered in this way. The gap-closing phenomenon is accompanied by the gap-parity change which indicates the transition between a normal 2D insulator and a 2D TI. In the case of $Sb_2Te_3$, the intrinsic gaps of 1, 2 and 3 QLs are measured to be around 700, 230 and 60 meV. As predicted theoretically, the gaps behave in an oscillatory manner with thicknesses[10]. The 3-QL $Sb_2Te_3$ film with a small gap here prove to be a tunable system in the presence of twist angles and possible strain. The possible gap-closing here indicates possible topological phase transition in the 3-QL film, regardless of whether the intrinsic film belongs to the non-trivial case or not. The unambiguous attribution of the gapless features shown in figure 4 to the gap-closing calls for further study.

In conclusion, we successfully grow twisted structures of 2-QL and 3-QL $Sb_2Te_3$ films by regulating MBE parameters and extremising the non-equilibrium growth mode.

The resulting twisted structures are mainly divided into two categories, one with anti-parallel alignment and the other with parallel or perpendicular alignment between the 2$^{nd}$ and 3$^{rd}$ QLs, respectively. The first case drive the system into a large-gap regime, while the second case retains the small gap of the 3-QL film, in which the gap can be modified and possible signature of gap-closing is presented here. Our approach sheds new light on the search of quantum spin Hall materials especially in the 3D TI systems.

**Acknowledgments**

We acknowledge the supporting from National Science Foundation and Ministry of Science and Technology of China (Grants No. 61804056). Y.P.J. acknowledges support from the National Thousand-Young-Talents Program.

**References**

[1]  M. Z. Hasan and C. L. Kane, Rev. Mod. Phys. **82**, 3045 (2010).
[2]  Y. L. Chen *et al.*, Science **325**, 178 (2009).
[3]  Y. Jiang *et al.*, Phys. Rev. Lett. **108**, 016401 (2012).
[4]  Y. Xia *et al.*, Nat. phys. **5**, 398 (2009).
[5]  X.-L. Qi and S.-C. Zhang, Rev. Mod. Phys. **83**, 1057 (2011).
[6]  H. Zhang, C.-X. Liu, X.-L. Qi, X. Dai, Z. Fang, and S.-C. Zhang, Nat. Phys. **5**, 438 (2009).
[7]  L. Fu, C. L. Kane, and E. J. Mele, Phys. Rev. Lett. **98**, 106803 (2007).
[8]  Y. Zhang *et al.*, Nat. Phys. **6**, 584 (2010).
[9]  W.-Y. Shan, H.-Z. Lu, and S.-Q. Shen, New J. Phys. **12**, 043048 (2010).
[10] C.-X. Liu, H. Zhang, B. Yan, X.-L. Qi, T. Frauenheim, X. Dai, Z. Fang, and S.-C. Zhang, Phy. Rev. B **81**, 041307 (2010).
[11] B. A. Bernevig, T. L. Hughes, and S.-C. Zhang, Science **314**, 1757 (2006).
[12] M. König, S. Wiedmann, C. Brüne, A. Roth, H. Buhmann, L. W. Molenkamp, X.-L. Qi, and S.-C. Zhang, Science **318**, 766 (2007).
[13] Y. Yang, Z. Xu, L. Sheng, R. Shen, B. G. Wang, and D. Y. Xing, EPL **100**, 27005 (2012).
[14] M. Kim, C. H. Kim, H.-S. Kim, and J. Ihm, Proc. Natl. Acad. Sci. U.S.A **109**, 671 (2012).
[15] S. Murakami, S. Iso, Y. Avishai, M. Onoda, and N. Nagaosa, Phys. Rev. B **76**, 205304 (2007).
[16] Y. Cao, V. Fatemi, S. Fang, K. Watanabe, T. Taniguchi, E. Kaxiras, and P. Jarillo-Herrero, Nature **556**, 43 (2018).
[17] Y. Jiang *et al.*, Phys. Rev. Lett. **108**, 066809 (2012).
[18] G. Wang et al., Adv. Mater. **23**, 2929 (2011).
[19] C. L. Song et al., Appl. Phys. Lett. **97**, 143118 (2010).
[20] G. Zhang et al., Appl. Phys. Lett. **95**, 053114 (2009).
[21] Y.Y. Li et al., Adv. Mater. **22**, 4002 (2010).
[22] T. Zhang, J. Ha, N. Levy, Y. Kuk, and J. Stroscio, Phys. Rev. Lett. **111**, 056803 (2013).

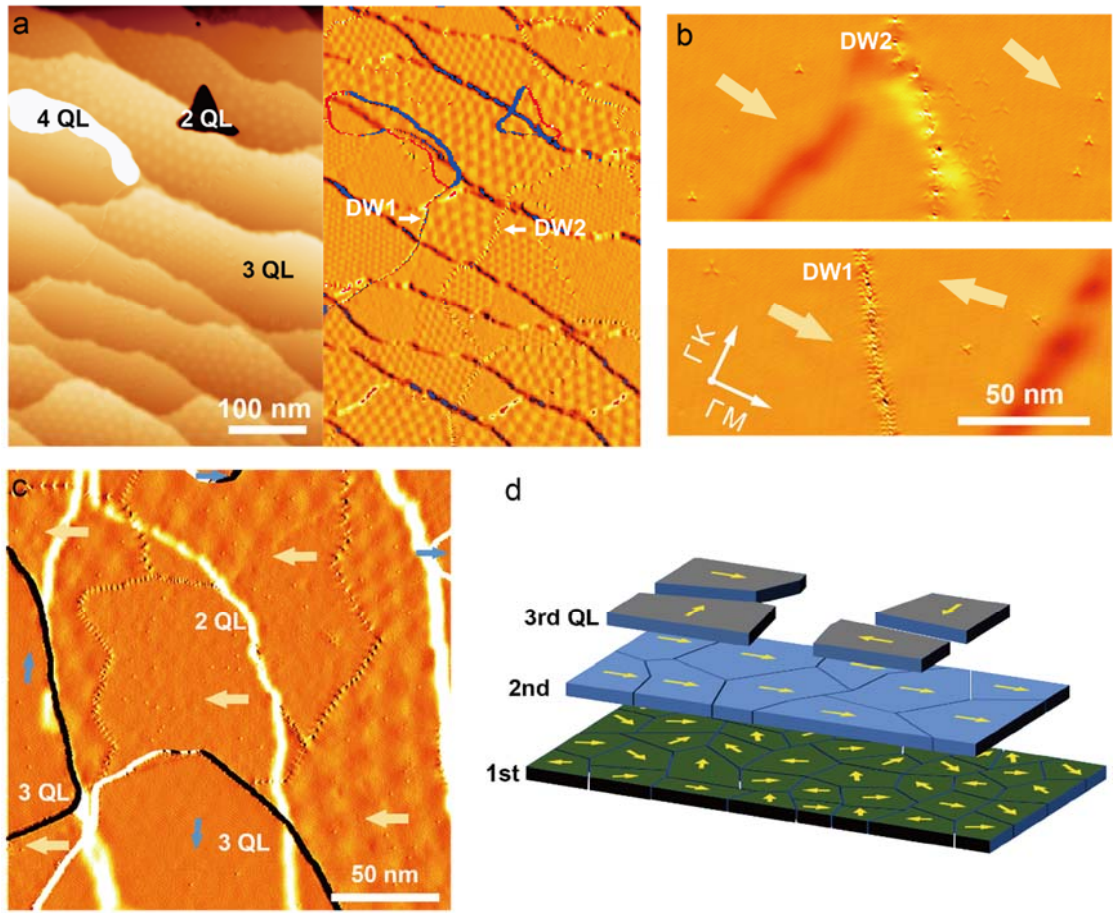

Figure 1. a. STM image and the corresponding differential image (3.0 V, 10 pA) of a 3-QL $Sb_2Te_3$ film. The substrate is STO(111). b. Derivative STM images of an $Sb_2Te_3$ film showing two different domain boundaries DW1 and DW2. The arrows indicate the $\overline{\Gamma}-\overline{M}$ directions of the $Sb_2Te_3$ crystal structure projected onto the surface brillouin zone. c. A derivative STM image of an $Sb_2Te_3$ film (2.6 QLs) showing the crystal orientations of domains in 2nd and 3rd QLs. d. A schematic layout of the domain structures in the $Sb_2Te_3$ film of 2.6 QLs.

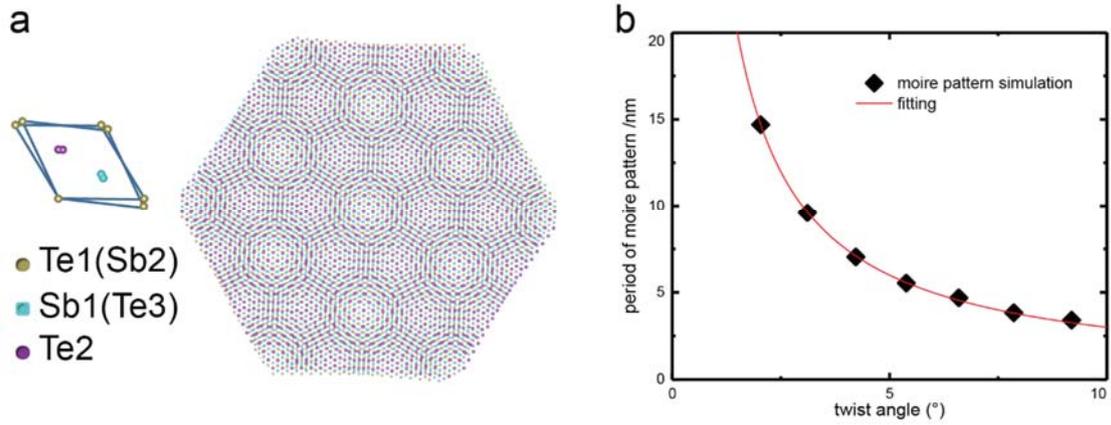

Figure 2. a. Moire pattern simulation (~5° mismatch). The insert shows the surface primary cell of 1-QL $Sb_2Te_3$. The brown, cyan, purple dots show the positions of Te1 and Sb2, Sb1 and Te3, Te2 projected onto the surface, respectively. The atoms are numbered from the surface. b. Simulated moire pattern periods for a 2-QL $Sb_2Te_3$ film with different twist angels. The red curve is the fitting to the data.

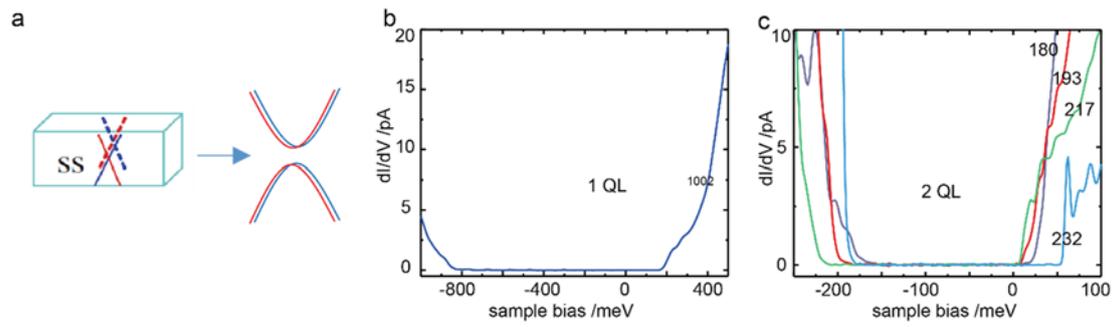

Figure 3. a. A simple diagram showing the evolution of top and bottom surface states. A hybridization gap opens in the surface states when the film is thin enough. b. A typical STS of 1-QL $Sb_2Te_3$ films. The numbers indicate the gap values. c. STS of 2-QL $Sb_2Te_3$ films showing different hybridization gap in different domains.

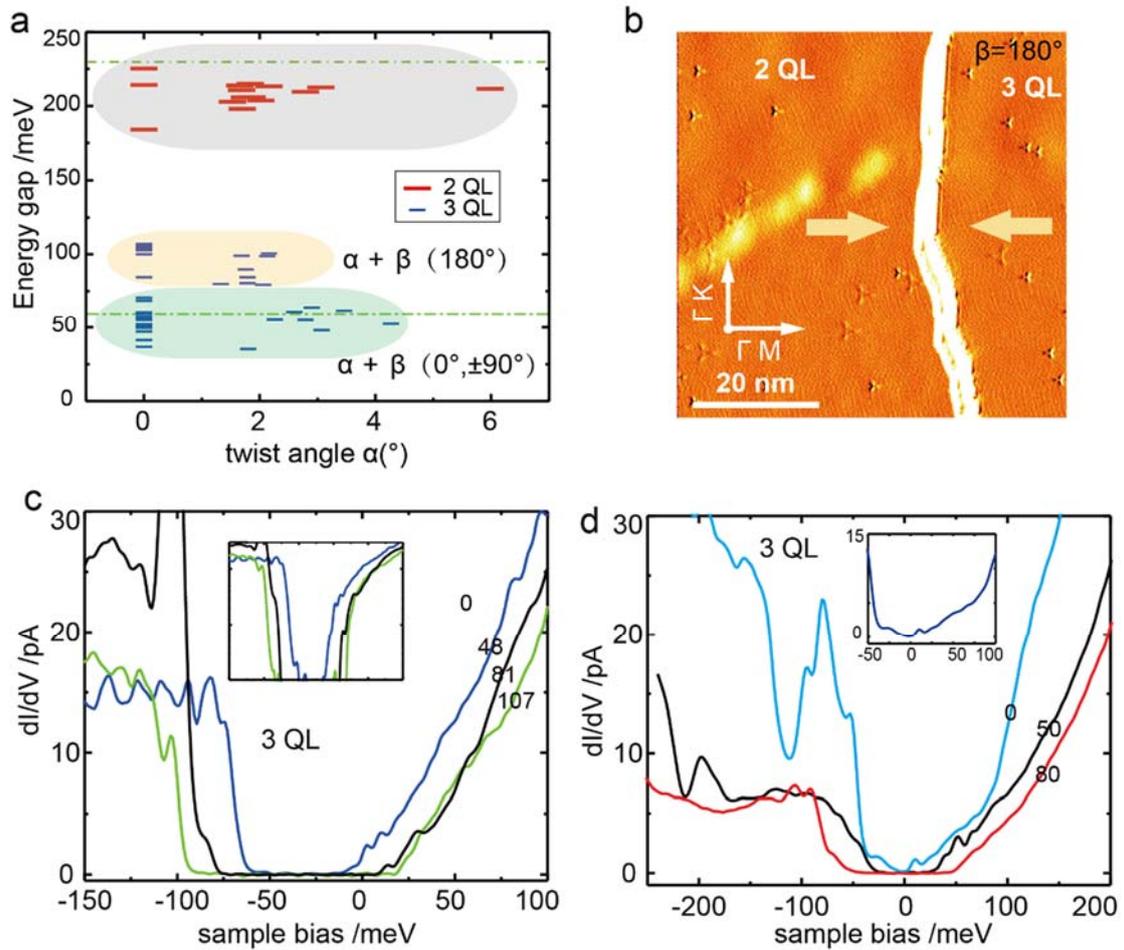

Figure 4. a. Statistics of gap values in 2- and 3-QL films. The dashed lines indicate the gap values for 2- and 3-QL films grown on graphene substrates. b. A derivative STM image showing the flipped crystal orientations between 2$^{nd}$ and 3$^{rd}$ QLs. c. STS of 3-QL $Sb_2Te_3$ films showing different hybridization gap in different regions. The inserts are the corresponding log-scaled plots. d. Typical STS for 3-QL films with different gaps in a larger energy scale. The insert shows the gapless-like spectrum (light blue) in a smaller energy range.